\documentclass[a4paper,11pt,dvips]{article}
\usepackage{graphicx}
\usepackage{amsmath}
\usepackage{amssymb}
\usepackage{latexsym}
\usepackage[colorlinks]{hyperref}
\usepackage{color}
\usepackage{float}
\usepackage{cite}
\usepackage{makeidx}
\usepackage{colortbl}
\newcommand{\bra}{\begin{array}}
\newcommand{\era}{\end{array}}
\newcommand{\beq}{\begin{equation}}
\newcommand{\eeq}{\end{equation}}
\newcommand{\bqr}{\begin{eqnarray}}
\newcommand{\eqr}{\end{eqnarray}}

\def\BC{\bb C}
\def\_\BC{\bbi C}

\def\no2 {{\textstyle{n\over 2}}}

\newcommand{\ga}{\gamma}

\newcommand{\lb}{\label}

\begin{document}

\begin{titlepage}
\setcounter{page}{1}
\renewcommand{\thefootnote}{\fnsymbol{footnote}}

\begin{flushright}
%ucd-tpg:****.**\\
%arXiv:yymm.xxxx
\end{flushright}

\vspace{5mm}
\begin{center}

{\Large \bf {Effect of Magnetic Field on Goos-H\"anchen Shifts \\ in Gaped  Graphene
Triangular Barrier}}

\vspace{5mm} {\bf Miloud Mekkaoui}$^{a}$,
 {\bf Ahmed Jellal\footnote{\sf
a.jellal@ucd.ac.ma}}$^{a,b}$ and {\bf Hocine Bahlouli}$^{b,c}$

\vspace{5mm}

{$^{a}$\em Laboratory of Theoretical Physics,  %Department of Physics,
Faculty of Sciences, Choua\"ib Doukkali University},\\
{\em PO Box 20, 24000 El Jadida, Morocco}

{$^b$\em Saudi Center for Theoretical Physics, Dhahran, Saudi Arabia}

{$^c$\em Physics Department,  King Fahd University
of Petroleum $\&$ Minerals,\\
Dhahran 31261, Saudi Arabia}

%\vspace{30mm}

\vspace{3cm}

\begin{abstract}

We study the effect of a magnetic field on
Goos-H\"anchen shifts  in gaped graphene
subjected to   a double triangular barrier.
Solving the wave equation separately in each region composing
our system and using the required boundary conditions, we then compute
explicitly the transmission probability for scattered fermions. These
wavefunctions are then used to derive
the Goos-H\"anchen  shifts
in terms of different physical parameters such as energy,
electrostatic potential strength and magnetic field. Our numerical results
show that the Goos-H\"anchen shifts are affected by the presence of the magnetic field
and depend on the geometrical structure
of the triangular barrier.

\end{abstract}
\end{center}

\vspace{3cm}


\end{titlepage}

%\newpage

%%%%%%%%%%%%%%%%%%%%%%%%%%%%%%%%%%%%%%%%%%%%%%%%%%
\section{Introduction}
%%%%%%%%%%%%%%%%%%%%%%%%%%%%%%%%%%%%%%%%%%%%%%%%%%%%%%

%\hspace{.33in}
Quantum and classical analogies between  phenomena occurring in
two different physical systems
can be at the origin of discovering new effects that are relevant in device applications,
%provide a route to find new effects in the devices,
and often help to understand both systems
better~\cite{Dragoman}. It is well known that a light beam totally
reflected from an interface between two dielectric media undergoes
lateral shift from the position predicted by geometrical
optics~\cite{Goos}. The close relationship between optics and
electronic results from the fact that the electrons behave as de
Broglie wave due to the ballistic transport properties of a highly
mobility two-dimensional electron gas created in semiconductor
heterostructures~\cite{Gaylord}.
The recent discovery of graphene~\cite{Wallace, Novoselov}
 added a new twist to this well established optical
analogy of ballistic electron transport. % and vice verse.

Graphene remains among the most fascinating and attractive
subject in  condensed matter physics. This is because of its
exotic physical properties and the apparent similarity of its
mathematical model to the one describing relativistic fermions in
two-dimensions. As a consequence of this relativistic-like
behavior particle could tunnel through very high barriers in
contrast to the conventional tunneling of non-relativistic
particles, an effect known in relativistic field theory as Klein
Tunneling. This tunneling effect has already been observed
experimentally~\cite{Stander} in graphene systems. There are
various ways for creating barrier structures in
graphene~\cite{Katsnelsonn, Sevin}, for instance it can be
done by applying a gate voltage, cutting the graphene sheet into
finite width to create  nanoribbons, using doping or through the
creation of a magnetic barrier. In the case of graphene, computation
of the transmission coefficient and the tunneling conductance were
already reported for electrostatic barriers~\cite{Sevin,
Masir, Dell'Anna, Mukhopadhyay}, magnetic
barriers~\cite{Dell'Anna, Choubabi, Mekkaoui}, potential
barrier~\cite{Jellal} and triangular barrier~\cite{HBahlouli}.

%\hspace{.33in}

 During the past few years there was a progress in studying the optical properties
 %of electron transport properties
 in graphene systems such as %. Among them, we cite
 the quantum version of
the  Goos-H\"anchen (GH) effect originating from the reflection of
particles from interfaces.  The GH effect was discovered by Hermann
Fritz Gustav Goos and Hilda H\"anchen~\cite{Goos} and
theoretically explained by Artman~\cite{Artmann} in the late of
1940s. Many works in various graphene-based nanostructures,
including single~\cite{Chen15}, double barrier~\cite{Song16}, and
superlattices~\cite{Chen18}, showed that the GH shifts can be
enhanced by the transmission resonances and controlled by varying
the electrostatic potential and induced gap~\cite{Chen15}. Similar
to observations of GH shifts in semiconductors, the GH shifts in graphene can also be
modulated by electric and magnetic barriers~\cite{Sharma19}, an analogous GH like shift
can also be observed in
atomic optics~\cite{Huang13}. It has been reported that the GH
shift plays an important role in the group velocity of
quasiparticles along interfaces of graphene p-n
junctions~\cite{Beenakker,Zhao11}.
{Experimentally 
%graphene is one of materials that has been intensively
%used as a platform to investigate the GH effect. For example, 
when graphene
is deposited  on dielectric materials it results in a profound effect on GH where
it can be either positive or negative with a complete electrostatic control \cite{R1,R2}.
Recently it has been shown that nonlinear surface plasmon resonance in graphene can provide
rigorous enhancement and control over GH effect \cite{R3,R4}}

Very recently, we have studied the GH shift exhibited by  Dirac
fermions in graphene scattered by triangular double barrier~\cite{MMekkaoui}.
We extend our former work~\cite{MMekkaoui} to include the effect of an external magnetic field
on the Goos-H\"anchen shifts in a gaped graphene triangular barrier.
 By
separating our system into five regions, we determine the
solutions of the energy spectrum in terms of different physical
quantities in each region. After matching the wavefunctions at different interfaces
of potential width, we determine the transmission probability
and subsequently the GH shifts. To acquire a better understanding of our
results, we plot the GH shifts for different values of  the physical parameters
characterizing our system. We also show that
the GH shifts can be
influenced by the applied magnetic field
and can be positive or negative according
to the values taken by the physical parameters. In addition,
interesting discussions
and comments will be reported in different occasions.

%\newpage
%\hspace{.33in}
The present paper is organized as follows. In section 2, we formulate our
model by setting the Hamiltonian system describing particles
scattered by a triangular double barrier whose intermediate zone is
subject to a mass term. We also obtain the spinor
solution corresponding to each regions composing our system in terms of
different physical parameters.  Using
the transfer matrix resulting from the boundary conditions,  we
determine the corresponding transmission probability in section 3.
In section 4, we derive the analytical form of the GH shifts and in section 5
%require to
%split the energy into three domains in order to calculate  the
%phase shift and GH shifts. In section 4, we numerically
we present
the main numerical results for the GH shifts and  transmission probability of
the particle beam transmitted through graphene  triangular double barrier.
Finally, in section 6 we conclude our work and summarize our main findings.

%%%%%%%%%%%%%%%%%%%%%%%%%%%%%%%%%%%%%%%%%%%%%%%%%%
\section{ Model of the system}
%%%%%%%%%%%%%%%%%%%%%%%%%%%%%%%%%%%%%%%%%%%%%%%%%%%%%%

\begin{figure}[!ht]
  \centering
  \includegraphics[width=14cm, height=7cm ]{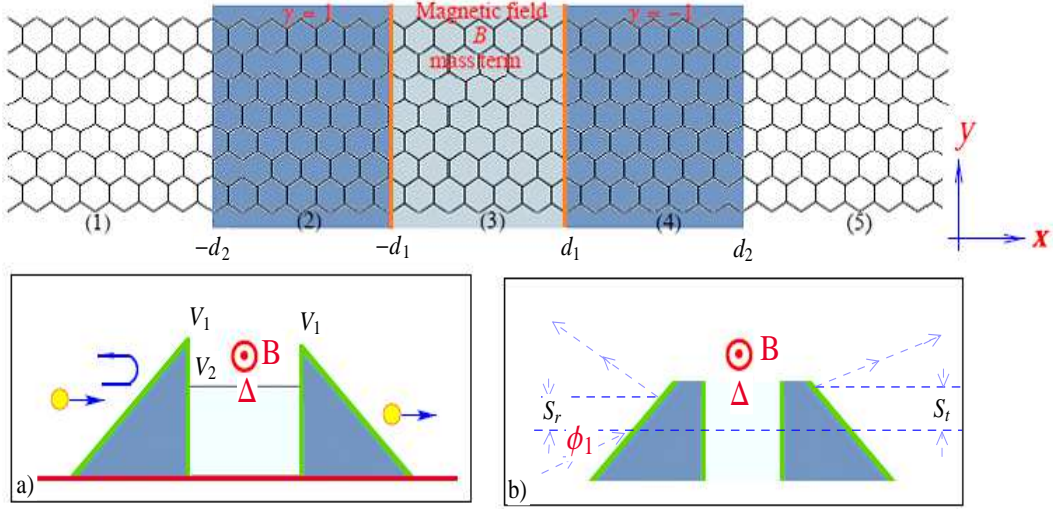}
  \caption{\sf Schematic plots of the models that we are considering, the light shaded area in the upper subplot
   represents Dirac fermions in inhomogeneous magnetic field through a graphene
triangular double barriers. (a): The shaded
area show smooth electric potentials with error function
distributions. (b): Describes the incident, reflected and
transmitted electron beams with  lateral shifts $S_r$ and
$S_t$.}\label{db.1}
\end{figure}

%\hspace{.33in}
We consider  massless Dirac particles with   energy $E$ and
 incident angle $\phi_1$ with respect to the incident $x$-direction
 of a gaped graphene  triangular double barrier.
%of height
%potential linear.
This system is a flat sheet of graphene subject
to a  triangular potential barrier along the $x$-direction while
particles are free in the $y$-direction. To ease our task let
us first describe the geometry of our system which is made of
five regions denoted by ${\sf j} = {\sf 1}, \cdots, {\sf 5}$. Each region is
characterized by its potential and interaction with external
sources. The barrier regions are formally described by the
Dirac-like Hamiltonian
\begin{equation}\lb{Ham1}
H=v_{F}
{\boldsymbol{\sigma}}\cdot\left(\textbf{p}+\frac{e}{c}\textbf{A}(x,y)\right)+
V(x){\mathbb I}_{2}+\Delta\sigma_{z}
\end{equation}
where ${v_{F}\approx 10^6 m/s}$  is the Fermi velocity,
${{\boldsymbol{\sigma}}=(\sigma_{x},\sigma_{y})}$ are the Pauli
matrices, $\textbf{p}=-i\hbar(\partial_{x},
\partial_{y})$, ${\mathbb I}_{2}$ the $2 \times 2$ unit matrix.
The vector potential will be chosen in the Landau gauge
$\textbf{A}(x,y) =
(0,A_{y}(x))$ with the magnetic field $\partial_{x}A_{y}(x)= B(x)$.
The parameter $\Delta = m v_{F}^2$ is
the energy gap owing to the sublattice symmetry breaking or can be
seen as  originating from
spin-orbit interaction $\Delta = \Delta_{so}$, and is confined to the region $|x|\leq d_1$
\begin{equation}
\Delta=t'\Theta\left(d_{1}^{2}-x^{2}\right)
\end{equation}
where  $\Theta$ is the Heaviside step function.
The double triangular barrier $V(x)$ is described by
the following
potential configuration
\begin{equation}
V(x)=V_{\sf j}=
\left\{%
\begin{array}{ll}
    (\gamma x+d_2)F, & \hbox{$d_{1}\leq |x|\leq d_{2}$} \\
    V_{2}, & \hbox{$ |x|\leq d_{1}$} \\
    0, & \hbox{otherwise} \\
\end{array}%
\right.
\end{equation}
where  $\gamma=1$ for $x\in [-d_2, -d_1]$,
$\gamma=-1$ for $x\in [d_1, d_2]$ and $F=\frac{V_1}{d_2-d_1}$,
 it is presented schematically in Figure \ref{db.1}.
We introduce a uniform perpendicular
magnetic field, along the $z$-direction, constrained to the well
region between the two barriers {(Figure \ref{db.1})} such that
\begin{equation}\label{eq04}
B(x)=B\Theta(d_{1}^{2}-x^{2})
\end{equation}
where $B$ is the strength of the magnetic field within the strip
located in the region $|x|>d_{1}$ and $B=0$ otherwise.
{We recall that  experimentally, different electrostatic profiles can be imposed on
the sample using different methods such as subjecting it to electrostatic gate or
changing the doping level as well as defects density. The triangular profile (Figure \ref{db.1})
%proposed in this article, 
or trapezoidal electrostatic barriers can be achieved
for example by imposing an inhomogeneous doping into the sample \cite{R5}.
On the other hand, the magnetic field profile imposed on the sample can be
provided externally using a magnetic strip.}

Choosing
the Landau gauge and imposing continuity of the vector potential
at the boundaries to avoid nonphysical effects, we end up with the
following vector potential
\begin{equation}
\qquad A_{y}(x)=A_{\sf j}=\frac{c}{el_{B}^{2}}\times\left\{%
\begin{array}{ll}
  -d_{1}, & \hbox{$x<-d_{2}$} \\
    x, & \hbox{$\mid x\mid<d_{1}$} \\
    d_{1}, & \hbox{$x\geq d_{2}$} \\
\end{array}%
\right.
\end{equation}
where the magnetic length is $l_{B}=\sqrt{1/B}$ in the unit
system $\hbar=c=e=1$. %Once we have defined the potential
%parameters relevant to all regions.
Recall that our system contains five regions denoted ${\sf j=1, \cdots, 5}$. The
left region (${\sf j=1}$) describes the incident electron beam
with energy $E=v_{F}\epsilon$ and  incident angle
$\phi_{1}$  where $v_{F}$ is the Fermi velocity. The right region
(${\sf j=5}$) describes the transmitted electron beam with
angle $\phi_{5}$.
The time-independent Dirac equation for the
spinor
$\psi_{\sf j}(x,y)=\left(\varphi_{\sf j}^{+},\varphi_{\sf j}^{-}\right)^{T}$
at energy $E=v_{F}\epsilon$  reads as
\begin{equation} \lb{eqh1}
\left[{\boldsymbol{\sigma}}\cdot\left(\textbf{p}+\frac{e}{c}\textbf{A}(x,y)\right)+
v_{\sf j} {\mathbb
I}_{2}+\mu\Theta\left(d_{1}^{2}-x^{2}\right)\sigma_{z}\right]\psi_{\sf j}(x,y)=\epsilon
\psi_{\sf j}(x,y)
\end{equation}
where we have defined $V_{\sf j}=v_{F}v_{\sf j}$ and $t'=v_{F}\mu$ and $F=v_{F}\varrho$.
 To proceed
further, we need to find the solutions of the corresponding Dirac
equation according to each region ${\sf j=1, \cdots, 5}$. Indeed,
for $x<-d_{2}$ (region 1):
\beq
\epsilon=
\left[p_{x1}^{2}+\left(p_{y}-\frac{1}{l_{B}^{2}}d_{1}\right)^{2}\right]^{\frac{1}{2}}
\eeq
\begin{eqnarray}
&& \psi_{1}(x,y)=\frac{1}{\sqrt{2}}\left(
\begin{array}{c}
1 \\
 z_{1}\end{array}\right)e^{\textbf{\emph{i}}(p_{x1} x + p_{y} y)}+r\frac{1}{\sqrt{2}}\left(
\begin{array}{c}
1 \\
 -z^{*}_{1}\end{array}\right)e^{\textbf{\emph{i}}(-p_{x1} x +p_{y}
 y)}
 \\
&&
z_{1}=s_{1}\frac{p_{x1}
+i\left(p_{y}-\frac{1}{l_{B}^{2}}d_{1}\right)}{\sqrt{p_{x1}^{2}
+\left(p_{y}-\frac{1}{l_{B}^{2}}d_{1}\right)^{2}}}
\end{eqnarray}
For $x > d_{2}$ (region 5):
\begin{eqnarray}
&& \epsilon=\left[p_{x5}^{2}+\left(p_{y}+\frac{1}{l_{B}^{2}}d_{1}\right)^{2}\right]^{\frac{1}{2}}\\
%\end{equation}
 %the solution is as follows:
%\begin{equation}
&&
\psi_{5}(x,y)=\frac{1}{\sqrt{2}}t\left(
\begin{array}{c}
1 \\
 z_{5}\end{array}\right)e^{\textbf{\emph{i}}(p_{x5} x +p_{y} y)},
 \qquad
z_{5}=s_{5}\frac{p_{x5}
+i\left(p_{y}+\frac{1}{l_{B}^{2}}d_{1}\right)}{\sqrt{p_{x1}^{2}
+\left(p_{y}+\frac{1}{l_{B}^{2}}d_{1}\right)^{2}}}
\end{eqnarray}

As far as  $|x|< d_{1}$ (region 3) is concerned, we first write down the corresponding
Hamiltonian in terms of  annihilation and creation operators. This can be obtained
from \eqref{Ham1}
\begin{equation}\label{eq 20}
H=v_{F}\left(%
\begin{array}{cc}
m^{+} & -i\frac{\sqrt{2}}{l_B} a^-\\
 i\frac{\sqrt{2}}{l_B} a^{+}  & m^{-} \\
\end{array}%
\right)
\end{equation}
where we have introduced
the shell operators
\beq
a^{\pm}=\frac{l_B}{\sqrt{2}}\left(\mp\partial_{x}+k_y+\frac{x}{l_{B}^{2}}\right)
\eeq
and the parameters
$m^{\pm}=
v_2\pm\mu$. We show that the involved operators obey the canonical commutation relation
$[a^-, a^{+}]={\mathbb I}$.
Note that, the energy gap $t'$ behaves like a mass
term in Dirac equation, which  this will affect the above
results and leads to interesting consequences on the transport
properties of our system.
We determine the eigenvalues and
eigenspinors of the Hamiltonian $H$ by considering the time
independent equation for the spinor $\psi_{3}(x, y)=(\psi_{3}^{+},
\psi_{3}^{-})^{T}$ using the fact that the transverse momentum
$p_{y}$ is conserved to write
$\psi_{3}(x, y)=e^{ip_{y}y} \varphi_{3}(x)$
with $\varphi_{3}(x)= (\varphi_{3}^+, \varphi_{3}^-)^{T}$.
Then the eigenvalue equation
\begin{equation}\label{eq 23}
H\left(%
\begin{array}{c}
  \varphi_{3}^+ \\
  \varphi_{3}^-\\
\end{array}%
\right)=\epsilon\left(%
\begin{array}{c}
  \varphi_{3}^+\\
  \varphi_{3}^-\\
\end{array}%
\right)
\end{equation}
 gives the coupled equations
\begin{eqnarray}\label{eq 25}
  &&m^{+}\varphi_{3}^{+}-i\frac{\sqrt{2}}{l_{B}}a^-\varphi_{3}^-=\epsilon\varphi_{3}^+\\
&&\label{eq 26}
  i\frac{\sqrt{2}}{l_{B}}a^{+}\varphi_{3}^+ +
  m^{-}\varphi_{3}^{-}=\epsilon\varphi_{3}^-.
\end{eqnarray}
Injecting \eqref{eq 26} into \eqref{eq 25} we end up with  a differential
equation of second order for
 $\varphi_{3}^{+}$
\begin{equation}
(\epsilon-m^{+})(\epsilon-m^{-})\varphi_{3}^{+}=\frac{2}{l_{B}^{2}}a^-
a^{+}\varphi_{3}^{+}.
\end{equation}
This is in fact an equation of the harmonic oscillator
and therefore we identify
$\varphi_{3}^{+}$ with its eigenstates
$|n-1\rangle$
corresponding to the eigenvalues
\begin{equation}
\epsilon-v_{2}=s_{3} k_{\eta}=s_{3}\frac{1}{l_{B}}\sqrt{(\mu
l_{B})^{2}+2n}
\end{equation}
where we have set $k_{\eta}=s_{3}(\epsilon-v_{2})$,
$s_{3}=\mbox{sign}(\epsilon-v_{2})$ correspond to positive and
negative energy solutions. The second spinor component can  be
derived from \eqref{eq 26} to obtain
\begin{equation}
\varphi_{3}^{-}=s_{3}i\sqrt{\frac{k_{\eta} l_{B}-s_{3} \mu
l_{B}}{k_{\eta} l_{B}+s_{3} \mu l_{B}}} \mid n\rangle.
\end{equation}
Introducing the parabolic cylinder functions
\beq
D_{n}(x)=2^{-\frac{n}{2}}e^{-\frac{x^{2}}{4}}H_{n}\left(\frac{x}{\sqrt{2}}\right)
\eeq
to express the solution in region {\sf 3} as
\begin{equation} \psi_{\sf
3}(x,y)=b_{1}\psi_{3}^{+}(x,y)+b_{2}\psi_{3}^{-}(x,y)
\end{equation}
with the two components
\begin{equation}
\psi_{3}^{\pm}(x, y)=\frac{1}{\sqrt{2}}\left(%
\begin{array}{c}
 \sqrt{\frac{k_{\eta} l_{B}+s_{3} \mu l_{B}}{k_{\eta} l_{B}}}
 D_{\left(\left(k_{\eta} l_{B}\right)^{2}-(\mu l_{B})^{2} \right)/2-1}
 \left(\pm \sqrt{2}\left(\frac{x}{l_{B}}+k_{y}l_{B}\right)\right) \\
  \pm i\frac{s_{3}\sqrt{2}}{\sqrt{k_{\eta} l_{B}\left(k_{\eta} l_{B}+s_{3} \mu l_{B}\right)}}
  D_{\left(\left(k_{\eta} l_{B}\right)^{2}-\left(\mu l_{B}\right)^{2}\right)/2}
 \left(\pm \sqrt{2}\left(\frac{x}{l_{B}}+k_{y}l_{B}\right)\right) \\
\end{array}%
\right)e^{ik_{y}y}
\end{equation}

In $d_{1}<|x|<d_{2}$ (regions {\sf 2} ($\ga=-1$) and {\sf 4} ($\ga=+1$)): the general
solution can be written in terms of the parabolic cylinder
function \cite{Abramowitz, Gonzalez, HBahlouli} as
\begin{equation}\lb{hii1}
 \chi_{\gamma}^{+}=c_{1}
 D_{\nu_\gamma-1}\left(Q_{\gamma}\right)+c_{2}
 D_{-\nu_\gamma}\left(-Q^{*}_{\gamma}\right)
\end{equation}
where we have set the parameters
$\nu_{\gamma}=\frac{i}{2\varrho}\left(k_{y}-\gamma\frac{d_{1}}{l_{B}^{2}}\right)^{2}$,
$\epsilon_{0}=\epsilon-v_{1}$ and made the change of variable
$
Q_{\gamma}(x)=\sqrt{\frac{2}{\varrho}}e^{i\pi/4}\left(\gamma
\varrho x+\epsilon_{0}\right)
$,
$c_{1}$ and $c_{2}$ are two
constants. The second component is given by
\begin{eqnarray}\lb{hii2}
\chi_{\gamma}^{-}&=&-c_{2}\frac{1}{k_{y}-\gamma\frac{d_{1}}{l_{B}^{2}}}\left[
2(\epsilon_{0}+\gamma \varrho x)
 D_{-\nu_\gamma}\left(-Q^{*}_{\gamma}\right)
+
 \sqrt{2\varrho}e^{i\pi/4}D_{-\nu_\gamma+1}\left(-Q^{*}_{\gamma}\right)\right]\nonumber\\
 &&
 -\frac{c_{1}}{k_{y}-\gamma\frac{d_{1}}{l_{B}^{2}}}\sqrt{2\varrho}e^{-i\pi/4}
 D_{\nu_\gamma-1}\left(Q_{\gamma}\right)
\end{eqnarray}
The components of the spinor solution
$\psi_{\sf m}(x,y) =\left(%
\begin{array}{c}
 \varphi_{\gamma}^{+}(x) \\
  \varphi_{\gamma}^{-}(x) \\
\end{array} \right)
e^{ik_{y}y}$
of the Dirac equation
\eqref{eqh1} in regions {\sf m=2} ($\gamma=-1$) and {\sf m=4} ($\gamma=+1$) can be obtained from
\eqref{hii1} and \eqref{hii2} by setting
\beq
\varphi_{\gamma}^{+}(x)=\chi_{\gamma}^{+}+i\chi_{\gamma}^{-}, \qquad
\varphi_{\gamma}^{-}(x)=\chi_{\gamma}^{+}-i\chi_{\gamma}^{-}
\eeq
and
therefore obtain
the eigenspinor
\begin{eqnarray}
 \psi_{\sf m}(x,y) &=& a_{{\sf m}-1}\left(%
\begin{array}{c}
 u^{+}_{\gamma}(x) \\
  u^{-}_{\gamma}(x) \\
\end{array}%
\right)e^{ik_{y}y}+a_{\sf m}\left(%
\begin{array}{c}
 v^{+}_{\gamma}(x) \\
 v^{-}_{\gamma}(x)\\
\end{array}%
\right)e^{ik_{y}y}
\end{eqnarray}
where  the functions
$u^{\pm}_{\gamma}(x)$ and $v^{\pm}_{\gamma}(x)$ are given by
\begin{eqnarray}
u^{\pm}_{\gamma}(x) &=&
 D_{\nu_{\gamma}-1}\left(Q_{\gamma}\right)\mp
 \frac{1}{k_{y}-\gamma\frac{d_{1}}{l_{B}^{2}}}\sqrt{2\varrho}e^{i\pi/4}D_{\nu_{\gamma}}\left(Q_{\gamma}\right)\\
%\end{eqnarray}
%\begin{eqnarray}
v^{\pm}_{\gamma}(x)&=&
 \pm\frac{1}{k_{y}-\gamma\frac{d_{1}}{l_{B}^{2}}}\sqrt{2\varrho}e^{-i\pi/4}D_{-\nu_{\gamma}+1}
 \left(-Q_{\gamma}^{*}\right)\nonumber\\
 &&
  \pm
 \frac{1}{k_{y}-\gamma\frac{d_{1}}{l_{B}^{2}}}\left(-2i\epsilon_{0}\pm
 \left(k_{y}-\gamma\frac{d_{1}}{l_{B}^{2}}\right)-\gamma2i \varrho x\right)D_{-\nu_{\gamma}}
 \left(-Q_{\gamma}^{*}\right).
\end{eqnarray}
%and  $j=2, 4$, $\gamma=\pm 1$ is labelling also both regions.
with $a_{\sf m}$ and $a_{{\sf m}-1}$ are four constants.
The above established results will be used in the next section to derive
%different issues
%related to
the transmission and refection amplitudes.

%%%%%%%%%%%%%%%%%%%%%%%%%%%%%%%%%%%%%%%%%%%%%%%%%%
\section{ Transmission amplitude} % and Goos-H\"anchen shifts}
%%%%%%%%%%%%%%%%%%%%%%%%%%%%%%%%%%%%%%%%%%%%%%%%%%%%%%

 %As usual
 The coefficients $(a_1,a_2,a_3,a_4,b_1,b_2,r,t)$ can be
determined using the boundary conditions, continuity of the
eigenspinors at each interface. Based on different considerations,
we study the interesting properties of our system in terms of the
corresponding transmission probability. Before doing so, let us
simplify our writing using the following shorthand notation
\begin{eqnarray}
&&\vartheta_{\tau1}^{\pm}=D_{\left[(k_{\eta}l_{B})^{2}-(\mu
l_{B})^{2}\right]/2-1}
 \left[\pm \sqrt{2}\left(\frac{\tau d_{1}}{l_{B}}+k_{y}l_{B}\right)\right]\\
&& \zeta_{\tau1}^{\pm}= D_{\left[(k_{\eta}l_{B})^{2}-(\mu
l_{B})^{2}\right]/2}
  \left[\pm \sqrt{2}\left(\frac{\tau d_{1}}{l_{B}}+k_{y}l_{B}\right)\right]\\
  && f_{1}^{\pm}=\sqrt{\frac{k_{\eta}\pm
\mu}{k_{\eta}}}, \qquad
f_{2}^{\pm}=\frac{\sqrt{2/l_{B}^{2}}}{\sqrt{k_{\eta}(k_{\eta}\pm
\mu)}}\\
&& u^{\pm}_{\gamma}(\tau d_{1})=u^{\pm}_{\gamma, \tau1},\qquad
u^{\pm}_{\gamma}(\tau d_{2})=u^{\pm}_{\gamma, \tau 2}\\
&& v^{\pm}_{\gamma}(\tau d_{1})=v^{\pm}_{\gamma, \tau 1},\qquad
v^{\pm}_{\gamma}(\tau d_{2})=v^{\pm}_{\gamma, \tau 2}
\end{eqnarray}
where $\tau=\pm$. Now, requiring the continuity of the spinor
wavefunctions at each junction interface gives rise to a set of
equations. We prefer to express these relationships in terms of
$2\times 2$ transfer matrices between different regions,
$\mathcal{M}_{jj+1}$. Then the full transfer matrix over the
whole triangular double barrier can be written %, in an obvious notation,
as
\begin{equation}\label{syst1}
\left(%
\begin{array}{c}
  1 \\
  r \\
\end{array}%
\right)=\prod_{\sf j=1}^{\sf 4}\mathcal{M}_{\sf j j+1}\left(%
\begin{array}{c}
  t \\
  0 \\
\end{array}%
\right)=\mathcal{M}\left(%
\begin{array}{c}
  t \\
  0 \\
\end{array}%
\right)
\end{equation}
which  %where the total transfer matrix
is the product of four transfer
matrices that couple the wave function in the ${\sf j}$-th region to the
wave function in the $({\sf j} + 1)$-th region %, such as
\beq
\mathcal{M}=\mathcal{M}_{12}\cdot \mathcal{M}_{2
3}\cdot \mathcal{M}_{34}\cdot\mathcal{ M}_{45}
\eeq
and are given by
%which take the form
%These are given by
\begin{eqnarray}
&& \mathcal{M}=\left(%
\begin{array}{cc}
  \tilde{m}_{11} & \tilde{m}_{12} \\
  \tilde{m}_{21} & \tilde{m}_{22} \\
\end{array}%
\right)\\
%\end{equation}
%\begin{equation}
&& \mathcal{M}_{12}=\left(%
\begin{array}{cc}
   e^{-\textbf{\emph{i}}p_{x1} d_{2}} &e^{\textbf{\emph{i}}p_{x1} d_{2}} \\
  z_{1}e^{-\textbf{\emph{i}}p_{x1} d_{2}} & -z^{\ast}_{1} e^{\textbf{\emph{i}}p_{x1} d_{2}} \\
\end{array}%
\right)^{-1}\left(%
\begin{array}{cc}
u_{1,-2}^{+} &  v_{1,-2}^{+}\\
 u_{1,-2}^{-} & v_{1,-2}^{-}\\
\end{array}%
\right)\\
%\end{equation}
%\begin{equation}
&& \mathcal{M}_{23}=\left(%
\begin{array}{cc}
 u_{1,-1}^{+} &  v_{1,-1}^{+}\\
u_{1,-1}^{-}& v_{1,-1}^{-}\\
\end{array}%
\right)^{-1}\left(%
\begin{array}{cc}
\vartheta_{1}^{+} &\vartheta_{1}^{-} \\
\zeta_{1}^{+} &\zeta_{1}^{-} \\
\end{array}%
\right)\\
%\end{equation}
%\begin{equation}
&& \mathcal{M}_{34}=\left(%
\begin{array}{cc}
 \vartheta_{-1}^{+} &\vartheta_{-1}^{-}\\
  \zeta_{-1}^{+} & \zeta_{-1}^{-} \\
\end{array}%
\right)^{-1}\left(%
\begin{array}{cc}
 u_{-1,1}^{+} &  v_{-1,1}^{+}\\
 u_{-1,1}^{-} & v_{-1,1}^{-} \\
\end{array}%
\right)\\
%\end{equation}
%\begin{equation}
&& \mathcal{M}_{45}=\left(%
\begin{array}{cc}
 u_{-1,2}^{+} &  v_{-1,2}^{+}\\
 u_{-1,2}^{-} & v_{-1,2}^{-}\\
\end{array}%
\right)^{-1}\left(%
\begin{array}{cc}
  e^{\textbf{\emph{i}}p_{x5} d_{2}} & e^{-\textbf{\emph{i}}p_{x5} d_{2}} \\
  z_{5} e^{\textbf{\emph{i}}p_{x5} d_{2}}  & -z_{5}^{\ast} e^{-\textbf{\emph{i}}p_{x5} d_{2}}  \\
\end{array}%
\right).
\end{eqnarray}
Using these we obtain the  transmission and reflection
amplitudes
\begin{equation}\label{eq 63}
 t=\frac{1}{\tilde{m}_{11}}, \qquad  r=\frac{\tilde{m}_{21}}{\tilde{m}_{11}}
\end{equation}
and
explicitly,  $t$ takes the form
 \beq
t=
\frac{e^{id_{2}\left(p_{x1}+p_{x5}\right)}\left(1+z_{5}^{2}\right)
\left(\vartheta_{1}^{-}\zeta_{1}^{+}+\vartheta_{1}^{+}\zeta_{1}^{-}
\right)}{f_{2}^{+}\left(f_{1}^{-}\mathcal{L}_{1}+if_{2}^{-}
\mathcal{L}_{2}\right)+f_{1}^{+}\left(f_{2}^{-}\mathcal{L}_{3}+if_{1}^{-}
\mathcal{L}_{4}\right)}\mathcal{D}
 \eeq
 where we have set %these quantities $\mathcal{D}$, $\mathcal{L}_{1}$, $\mathcal{L}_{2}$, $\mathcal{L}_{3}$ and $\mathcal{L}_{4}$ are defined by
 \begin{eqnarray}
\mathcal{D}&=&
\left(u_{-1,1}^{-}v_{-1,1}^{+}-u_{-1,1}^{+}v_{-1,1}^{-} \right)
\left(u_{1,-2}^{+}v_{1,-2}^{-}-u_{1,-2}^{-}v_{1,-2}^{+}
\right)\\
 \mathcal{L}_{1}&=&
\vartheta_{-1}^{-}\zeta_{1}^{+}\mathcal{F}\mathcal{G}
-\vartheta_{1}^{-}\zeta_{-1}^{+}\mathcal{K}\mathcal{J}\\
\mathcal{L}_{2}&=&\left(\zeta_{1}^{+}\zeta_{-1}^{-}-\zeta_{1}^{-}\zeta_{-1}^{+}\right)\mathcal{F}\mathcal{J}\\
 \mathcal{L}_{3}&=&
\vartheta_{-1}^{+}\zeta_{1}^{-}\mathcal{F}\mathcal{G}-\vartheta_{1}^{+}\zeta_{-1}^{-}\mathcal{K}\mathcal{J}\\
\mathcal{L}_{4}&=&=\left(\vartheta_{1}^{+}\vartheta_{-1}^{-}-\vartheta_{1}^{-}\vartheta_{-1}^{+}\right)\mathcal{K}\mathcal{G}
\end{eqnarray}
with the quantities
\begin{eqnarray}
\mathcal{F}&=&\left[u_{1,-1}^{+}v_{1,-2}^{-}-u_{1,-2}^{-}v_{1,-1}^{+}-
z_{1}\left(u_{1,-1}^{+}v_{1,-2}^{+}-u_{1,-2}^{+}v_{1,-1}^{+}\right)\right]\\
\mathcal{G}&=&\left[u_{-1,1}^{-}v_{-1,2}^{+}-u_{-1,2}^{+}v_{-1,1}^{-}
+z_{5}\left(u_{-1,1}^{-}v_{-1,2}^{-}-u_{-1,2}^{-}v_{-1,1}^{-}\right)\right]\\
\mathcal{K}&=&\left[u_{1,-1}^{-}v_{1,-2}^{-}-u_{1,-2}^{-}v_{1,-1}^{-}-z_{1}
\left(u_{1,-1}^{-}v_{1,-2}^{+}-u_{1,-2}^{+}v_{1,-1}^{-}\right)\right]\\
\mathcal{J}&=&\left[u_{-1,1}^{+}v_{-1,2}^{+}-u_{-1,2}^{+}v_{-1,1}^{+}+z_{5}
\left(u_{-1,1}^{+}v_{-1,2}^{-}-u_{-1,2}^{-}v_{-1,1}^{+}\right)\right]
\end{eqnarray}
We can also write \eqref{eq 63} % the transmission  and reflection amplitudes in
 in complex notation as
\begin{equation}
 t= \rho_{t} e^{i\varphi_{t}}, \qquad
r = \rho_{r} e^{i\varphi_{r}}
\end{equation}
 where the phase of the transmitted $\varphi_{t}$ and reflected $\varphi_{r}$
 amplitudes are given by
%\begin{equation}
 %    \varphi_{t} = \arctan\left(\frac{\mbox{Im} [t]}{\mbox{Re} [t]}\right),
  %   \qquad \varphi_{r} = \arctan\left(\frac{\mbox{Im} [r]}{\mbox{Re} [r]}\right)
 %\end{equation}
 %with the quantities
%\begin{eqnarray}
%&&\mbox{Re} [t]=\frac{t+t^{\ast}}{2},\qquad
%\mbox{Im} [t]=\frac{t-t^{\ast}}{2i}\\
%&&
%\mbox{Re} [r]=\frac{r+r^{\ast}}{2},\qquad
 %\mbox{Im} [r]=\frac{r-r^{\ast}}{2i}
%\end{eqnarray}
%which are giving rise
% Now we are ready for the
%computation of the phase shift of the transmitted
%$\varphi_{t_\eta}$ and the reflected $\varphi_{r_\eta}$. which
%allows us to express the
 %phase shift is given by
 \begin{equation}
     \varphi_{t} = \arctan\left(i\frac{t^{\ast}-t}{t^{\ast}+t}\right),
     \qquad \varphi_{r} = \arctan\left(i\frac{r^{\ast}-r}{r^{\ast}+r}\right).
 \end{equation}

Actually what we exactly need are the transmission and
reflection  probabilities, which can be obtained by introducing
the electric current density $J$ corresponding to our system. Then
from the previous Hamiltonian, we derive
the incident, reflected and transmitted current
%the following quantities
\begin{eqnarray}
&& J_{\sf {inc}}=  e\upsilon_{F}(\psi_{1}^{+})^{\dagger}\sigma
_{x}\psi_{1}^{+}\\
%\end{equation}
%\begin{equation}
 &&
 J_{\sf {ref}}= e\upsilon_{F} (\psi_{1}^{-})^{\dagger}\sigma _{x}\psi_{1}^{-}\\
%\end{equation}
%\begin{equation}
&&
 J_{\sf {tra}}= e\upsilon_{F}\psi_{5}^{\dagger}\sigma _{x}\psi_{5}.
\end{eqnarray}
These
can be used to write the transmission and reflection probabilities as
\begin{equation}
  T= \frac{p_{x5}}{p_{x1}}\left(\mbox{Im}[t]^{2}+\mbox{Re}[t]^{2}\right), \qquad
  R=\mbox{Im}[r]^{2}+\mbox{Re}[r]^{2}.
\end{equation}
These will be numerically computed by choosing different values
of the parameters characterizing the present system. In fact, we will present
different plots to underline and understand the basic properties of our system.

%%%%%%%%%%%%%%%%%%%%%%%%%%%%%%%%%%%%%%%%%%%%%%%%%%%%%%%%%
\section{The Goos-H\"anchen shifts}
%%%%%%%%%%%%%%%%%%%%%%%%%%%%%%%%%%%%%%%%%%%%%%%%%%%%%%%%%

The Goos-H\"anchen   shifts in graphene can be analyzed by considering an
incident, reflected and transmitted beams around some transverse
wave vector $k_y = k_{y_0}$ and the angle of incidence
$\phi_{1}(k_{y_{0}})\in \left[0, \frac{\pi}{2}\right]$, denoted by the
subscript $0$. These can be expressed in integral forms  as
\begin{eqnarray}
   \Psi_{i}(x,y) &=& \int_{-\infty}^{+\infty}dk_y\ f(k_y-k_{y_0})\ e^{i(k_{x1}(k_y)x+k_yy)}\left(
            \begin{array}{c}
              {1} \\
              {e^{i\phi_{1}(k_{y})}}
            \end{array}
          \right)\label{eq 79}\\
%\end{eqnarray}
%and the reflected one is thus expressed as
%\begin{eqnarray}
\Psi_{r}(x,y) &=& \int_{-\infty}^{+\infty}dk_y\ r(k_y)\
f(k_y-k_{y_0})\ e^{i(-k_{x1}(k_y)x+k_yy)}\left(
            \begin{array}{c}
              {1} \\
              {-e^{-i\phi_{1}(k_{y})}} \\
            \end{array}
          \right)\label{refl}
\end{eqnarray}
and
the reflection coefficient is $r(k_y)=|r|e^{i\varphi_{r}}$. This
fact is represented by writing the $x$-component of wave vector,
$k_{x1}$ as well as $\phi_{1}$  in terms of $k_{y}$, where
each spinor plane wave is a solution of \eqref{eqh1} and
$f(k_y-k_{y_0})$ is the angular spectral distribution. We can
approximate the $k_{y}$-dependent terms by a Taylor expansion
around $k_{y}$, retaining only the first order term to get
\begin{eqnarray}
 &&
\phi_{1}(k_{y})\approx
\phi_{1}(k_{y_{0}})+\frac{\partial\phi_{1}}{\partial
k_{y}}|_{k_{y_{0}}}(k_{y}-k_{y_{0}})\\
&&
k_{x1}(k_{y})\approx k_{x1}(k_{y_{0}})+\frac{\partial
k_{x1}}{\partial k_{y}}|_{k_{y_{0}}}(k_{y}-k_{y_{0}}).
\end{eqnarray}
Finally, the transmitted beams are
\begin{eqnarray}
\Psi_{t}(x,y) &=& \int_{-\infty}^{+\infty}dk_y\ t(k_y)\
f(k_y-k_{y_0})\ e^{i(k_{x1}(k_y)x+k_yy)}\left(
            \begin{array}{c}
              {1} \\
              {e^{i\phi_{1}(k_{y})}} \\
            \end{array}
          \right)\label{trans}
\end{eqnarray}
 and the transmission coefficient is
$t(k_y)=|t|e^{i\varphi_{t}}$.
%coefficients will be calculated
%through the use of boundary conditions.

The stationary-phase
approximation indicates that  the GH shifts are equal to the
negative gradient of transmission phase with respect to $k_y$.
 To  calculate  the  GH  shifts  of  the
transmitted beam through our system, according to the stationary
phase method \cite{Bohm}, we adopt the definition \cite{Chen15,
AAJellal, AAAJellal}
 \begin{equation}
        S_{t}=- \frac{\partial \varphi_{t}}{\partial
        k_{y0}}, \qquad S_{r}=- \frac{\partial \varphi_{r}}{\partial
        k_{y0}}.\label{eq 80}
 \end{equation}
Assuming a finite-width beam with the Gaussian shape,
$f(k_y-k_{y0}) = w_{y} \exp\left[-(w^{2}_{ y}/2)(k_{y}-k_{y0})^{2}\right]$,
around $k_{y0}$, where $w_{y} = w \sec\phi_{1}$, and $w$ is the
half beam width at waist, we can evaluate the Gaussian integral to
obtain the spatial profile of the incident beam, by expanding
$\phi_{1}$ and $k_{x1}$ to first order around $k_{y0}$ when
satisfying the condition
\beq
\delta\phi_1= \lambda_{F} /(\pi w)\ll 1
\eeq
where $\lambda_{F}$ is Fermi wavelength. Comparison of the incident
and transmitted beams suggests that the displacements
$\sigma_{\pm}$ of up and down spinor components are both equal to
$\partial\varphi_t/\partial k_{y0}$ and the average displacement is
\beq
S_{t} = \frac{1}{2}(\sigma^{+}+ \sigma^{-}) = -\frac{\partial\varphi_t}
{\partial k_{y0}}.
\eeq
It should be noted that when the above-mentioned condition
%$\delta\phi= \lambda_{F} /(\pi w)\ll 1$
is satisfied, that is, the
stationary phase method is valid \cite{Chen15}, the definition
\eqref{eq 80} can be applicable to any finite-width beam, not
necessarily a Gaussian-shaped beam.

 %%%%%%%%%%%%%%%%%%%%%%%%%%%%%%%%%%%%%%%%%%%%%%%%%%
\section{ Discussion of numerical simulations}
%%%%%%%%%%%%%%%%%%%%%%%%%%%%%%%%%%%%%%%%%%%%%%%%%%%%%%

The numerical results for the GH shifts
of Dirac electrons in graphene
scattered by a triangular double barrier potential  under a
uniform vertical magnetic field are now presented. In fact, we numerically
evaluate the GH shifts  in transmission $S_{t}$ and in reflection $S_{r}$ regions, respectively,
as a function of structural parameters of our system, including
the energy $\epsilon l_B$, the
$y$-direction wave vector $k_{y}l_B$, the energy gap $\mu l_B$,
the barriers widths $d_1/l_B$ and $d_2/l_B$, the strength of
potential barriers  $v_{1}l_B$ and $v_{2}l_B$. First we
present in Figures \ref{GHfig2} and  \ref{GHfig3} the GH
shifts in the transmission region a) and the corresponding transmission
probabilities b) as  function of the incident energy $\epsilon
l_B$. We have chosen the parameters ($v_1l_B= 10$, $v_2l_B = 20$)
in Figure \ref{GHfig2} and ($v_1l_B= 20$, $v_2l_B= 10$) in Figure
\ref{GHfig3}, with $k_{y} l_{B}=1$ and three different values of
 the barrier width  $\frac{d_{1}}{l_{B}}=0.2$,
 $\frac{d_{1}}{l_{B}}=0.5$, $\frac{d_{1}}{l_{B}}=1.4$ corresponding
to red, green and blue colors, respectively.

\begin{figure}[!ht]
\centering
\includegraphics[width=8cm, height=5cm]{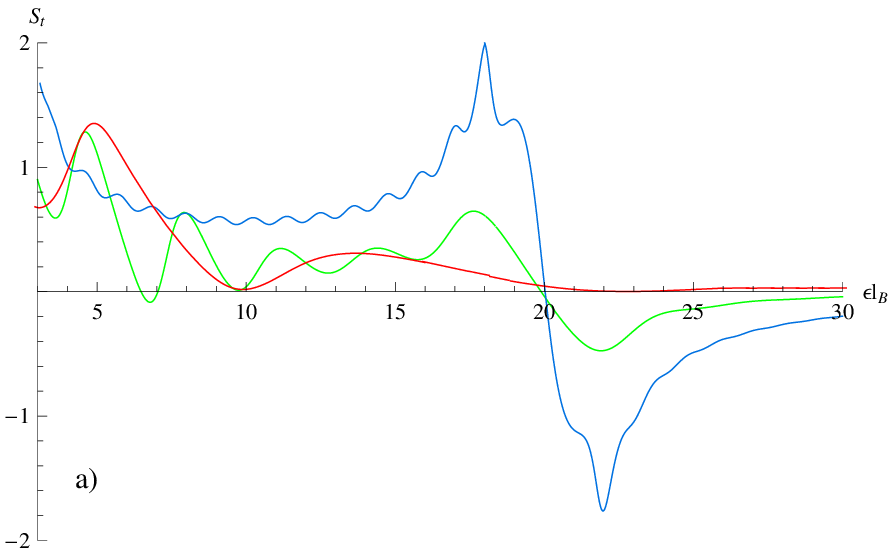}\ \ \ \
\includegraphics[width=8cm, height=5cm]{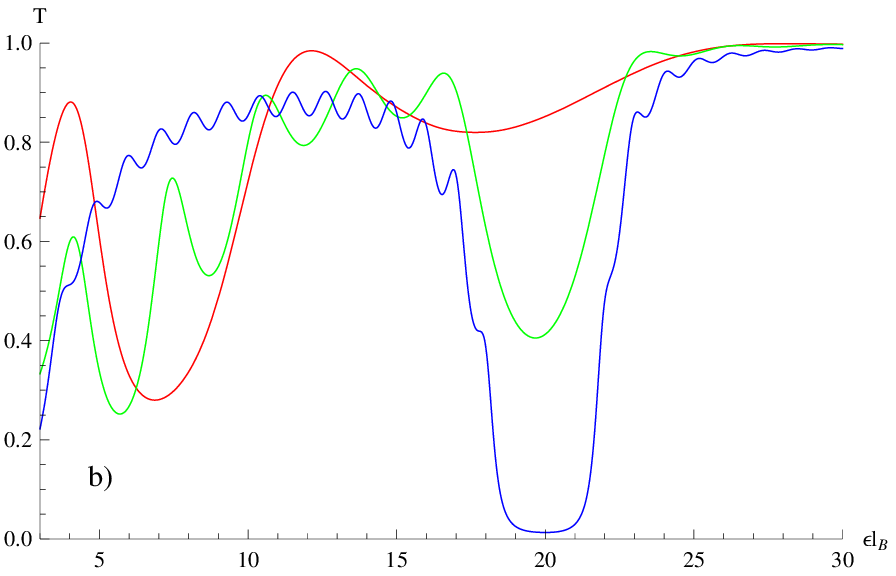}\\
%\end{figure}
%\begin{center}
 \caption{\sf{(Color online) a): The GH shifts in transmission $S_t$  and  b):
 the transmission probability $T$ versus the incident
energy $\epsilon l_{B}$ with $\frac{d_{1}}{l_{B}}=0.2$
red color, $\frac{d_{1}}{l_{B}}=0.5$ green color,
$\frac{d_{1}}{l_{B}}=1.4$ blue color,
 $\frac{d_{2}}{l_{B}}=1.5$, $\mu l_{B}=0$, $k_{y} l_{B}=1$ and $(v_{1}l_{B}=10, v_{2} l_{B}=20)$.}}\lb{GHfig2}
\end{figure}\begin{figure}[!ht]
\centering
\includegraphics[width=8cm, height=5cm]{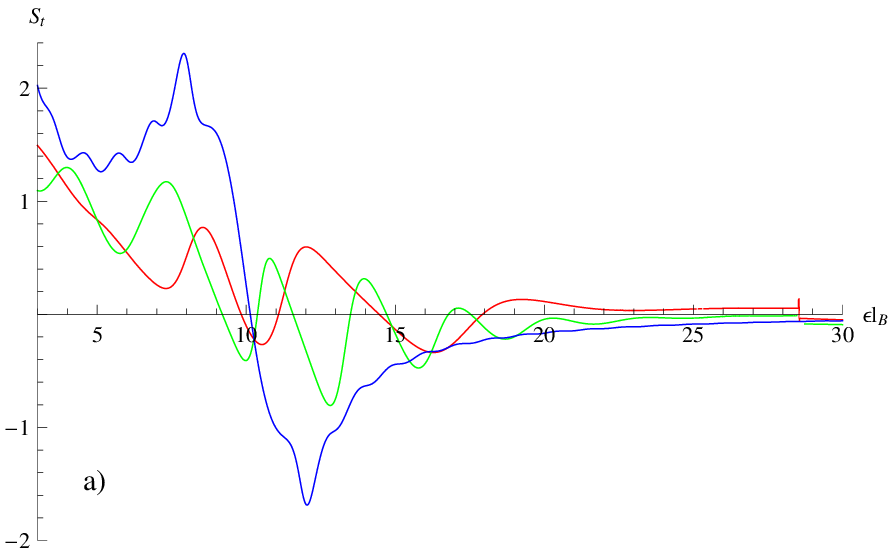}\ \ \ \
\includegraphics[width=8cm, height=5cm]{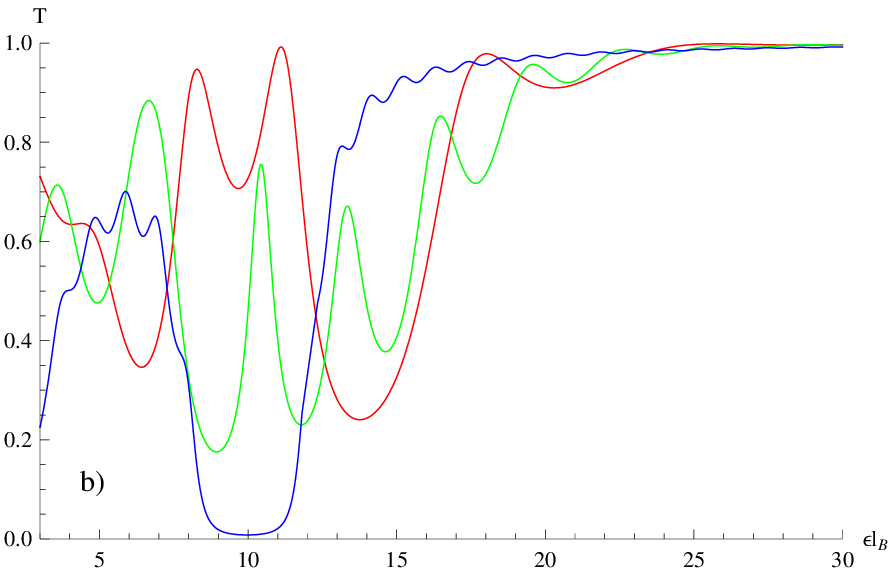}\\
%\end{figure}
%\begin{center}
 \caption{\sf{(Color online) a): The GH shifts in transmission $S_t$  and  b): the transmission probability $T$
 versus
the incident energy $\epsilon l_{B}$ with $\frac{d_{1}}{l_{B}}=0.2$
red color, $\frac{d_{1}}{l_{B}}=0.5$  green color,
$\frac{d_{1}}{l_{B}}=1.4$ blue color,
 $\frac{d_{2}}{l_{B}}=1.5$, $\mu l_{B}=0$, $k_{y} l_{B}=1$ and
 $(v_{1}l_{B}=20, v_{2} l_{B}=10)$.}}\lb{GHfig3}
\end{figure}
It is shown that the GH shifts $S_t$ are closely related to the
transmission probabilities. For the sake of simplicity, we will
find the explicit expressions at zero-gap $\mu l_{B}=0$. One can
notice that, at the Dirac points $\epsilon l_B=v_2l_B$, the GH
shifts change their sign and behave differently. % as compared.
This
change in sign of the GH shifts show clearly that they are
strongly dependent on the barrier heights. We also observe that the
GH shifts are negative and positive in Figures \ref{GHfig2} and
\ref{GHfig3}. Recall that, the Dirac points represent the zero modes
for Dirac operator \cite{Sharma19} and lead to the emergence of
new Dirac points, which have been discussed in different works
\cite{Bhattacharjee, Park1}. Such points separate the two regions
of positive and negative refraction. In the cases of $\epsilon
l_B<v_2l_B$ and $\epsilon l_B>v_2l_B$, the shifts are, respectively,
in the forward and backward directions, due to the fact that the
signs of group velocity are opposite. It is clearly seen that $S_t$ are
oscillating between negative and positive values around the
critical point $\epsilon l_B=v_2l_B$, in the interval when
$\epsilon l_B < v_1 l_B$  the usual high energy barrier
oscillations appear either in Figure \ref{GHfig2} or Figure \ref{GHfig3}.

\begin{figure}[!ht]
\centering
\includegraphics[width=8cm, height=5cm]{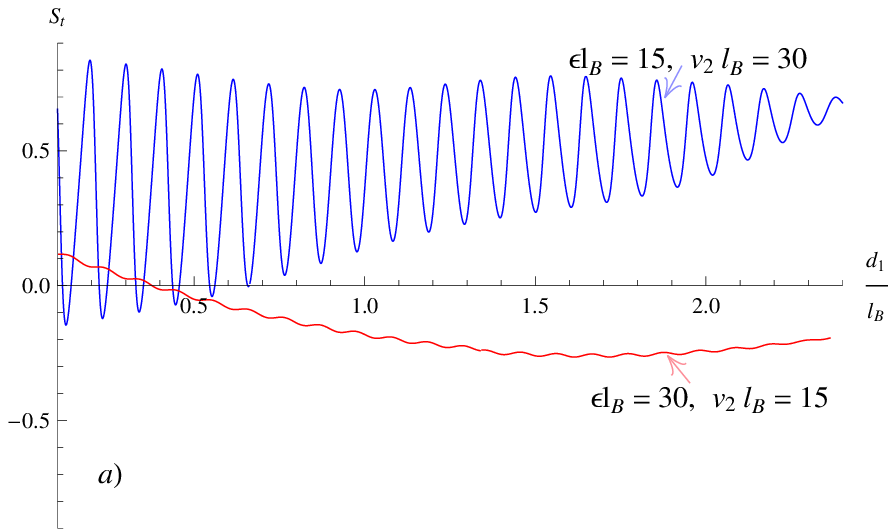}\ \ \ \
\includegraphics[width=8cm, height=5cm]{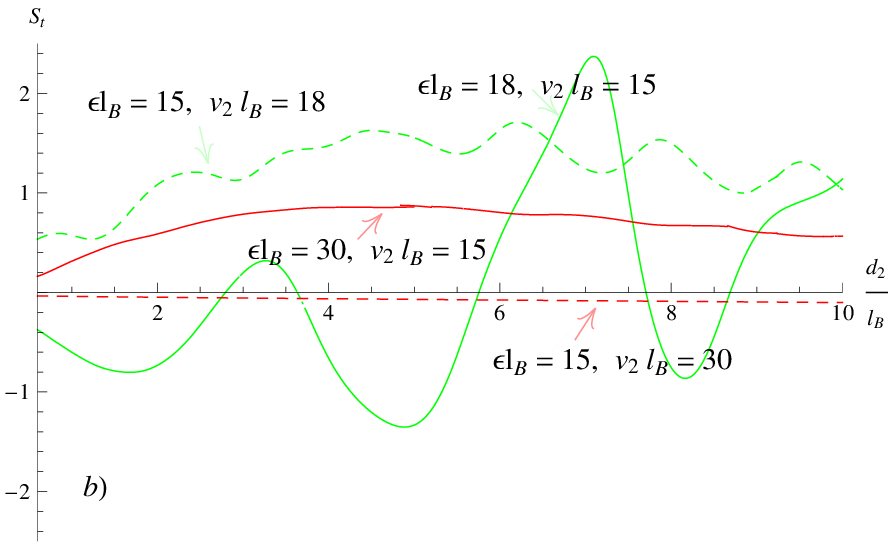}\\
%\end{figure}
%\begin{center}
 \caption{\sf{ (Color online)  The GH shifts in transmission $S_t$ versus  width
$\frac{d_{1}}{l_{B}}$ a) and width $\frac{d_{2}}{l_{B}}$ b) for triangular double barrier. a): For   $\mu l_{B}=0$,
 $\frac{d_{2}}{l_{B}}=2.5$, $k_{y} l_{B}=1$,
 $v_{1}l_{B}=25$, $(\epsilon=15, v_{2}=30\}$, $\{\epsilon=30, v_{2}=15)$.
 b): For   $\mu l_{B}=0$,
 $\frac{d_{1}}{l_{B}}=0.5$, $k_{y} l_{B}=1$,
 $v_{1}l_{B}=25$, $(\epsilon l_B=15, v_{2}l_B=18)$, $(\epsilon l_B=18, v_{2}l_B =15)$,
 $(\epsilon l_B=15, v_{2}l_B=30)$, $(\epsilon l_B=30, v_{2}l_B =15)$.}}\lb{GHfig4}
\end{figure}

 We further explore the effects of the triangular double barrier widths $d_1$ and $d_2$
 on the GH shifts in Figure \ref{GHfig4} for both cases
$\epsilon l_B<v_2l_B$ and $\epsilon
l_B>v_2l_B$.
Figure \ref{GHfig4}a) shows an interesting behavior of the GH shifts
in terms of the barrier width $d_1$
where  oscillations with different amplitudes appeared  for the
configuration $(\epsilon l_B=15, v_{2}l_B =30)$.
However such behavior completely changes when we reverse the choice of  parameters,
i.e. $\{\epsilon l_B=30, v_{2}l_B=15\}$ then $S_t$ crosses from positive to negative behaviors.
As far as the second barrier width $d_2$ is concerned,
from Figure \ref{GHfig4}b) we observe different oscillations by considering some values
of the couple
 $(\epsilon l_B, v_{2} l_B)$.

At this level,
we  turn to the discuss
the influence of the induced gap $\mu l_{B}$ in our system in the presence of
a triangular double
barrier and a magnetic field. Note that, the gap is introduced as shown in Figure
\ref{db.1} and therefore it affects the system energy according to
the solution of the energy spectrum obtained in region $3$.
  Figure \ref{GHfig5} shows that the GH
shifts in the propagating case can be enhanced by a gap opening at
the Dirac point. This has been performed by fixing the parameters
$\frac{d_1}{l_B}=1.1$, $\frac{d_2}{l_B} =1.5$, $v_1l_B=25$, $k_yl_B
=1$ and making different choices for the energy $\epsilon l_{B}$
and potential $v_2l_B$. Figure \ref{GHfig5}a) presents the GH
 shift in transmission $S_{t}$ and transmission probability as a function of
 energy gap $\mu l_B$. For the configuration $(\epsilon l_{B}=15, v_2l_B=23)$
 it is clear that one can still have
positive shifts (blue line) and  for the
configuration $(\epsilon l_{B}=23, v_2l_B=15)$ the GH shifts are
negative (red line). Note that for certain
energy gap $\mu l_B$, there is no transmission  possible
and therefore the GH
shifts in transmission $S_t$   vanish.
%It is shown
%that the GH shifts in transmission are closely related to the
%transmission probabilities.
As shown in Figure \ref{GHfig5}b), we
plot  the GH shifts in reflection $S_{r}$ and the reflection
probability  as a function of energy gap $\mu l_B$ and  found
that the GH shifts display sharp peaks inside the transmission
gap. It is clearly seen  that the GH shifts can be enhanced by a certain
gap opening. Indeed, by increasing the gap we observe that the gap
of transmission becomes broader,  changing the transmission
resonances and  the modulation of the GH shifts. Note that for
certain energy gap $\mu l_B$, there is  total reflection and therefore
the GH shifts in reflection $S_r$ does not vanish. In fact, under
the condition $\mu l_B>\mid\epsilon l_{B}-v_2l_B\mid$ every
incoming wave is reflected.
{In summary %Figure 5 shows both  GH shifts in reflected $S_r$  and transmitted $S_t$ beam 
%as a function of the energy gap. 
it is shown that ($S_r$, $R$) on one hand and ($S_t$, $T$) on the other 
hand are very much related in their structure.  Physically $S_t$ %the GH shift in the transmitted beam 
is detected in the transmission region by placing a detector in the outgoing region and far away 
from the incident region. On the 
contrary $S_r$ % the GH shift in the reflected beam 
is detected by placing a detector in the incident region far way 
from the transmission region.}

\begin{figure}[!ht]
\centering
\includegraphics[width=8cm, height=5cm]{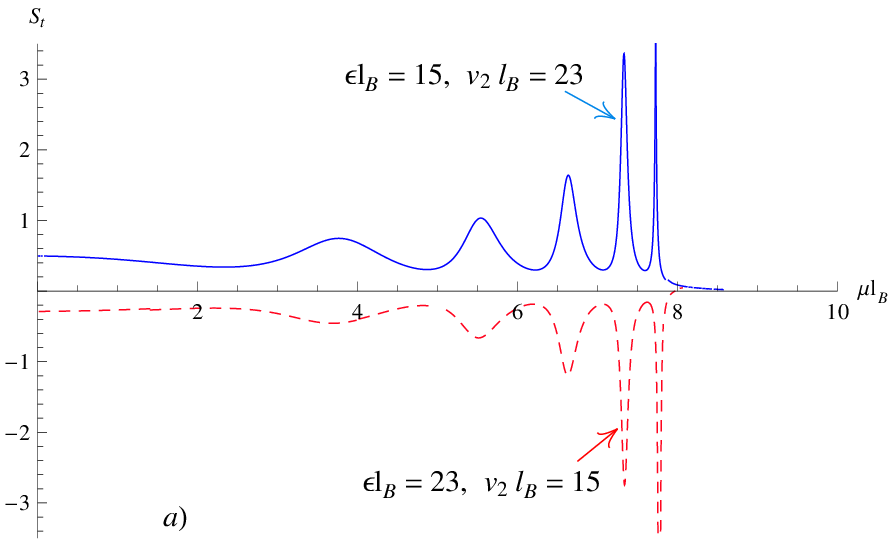}\ \ \ \
\includegraphics[width=8cm, height=5cm]{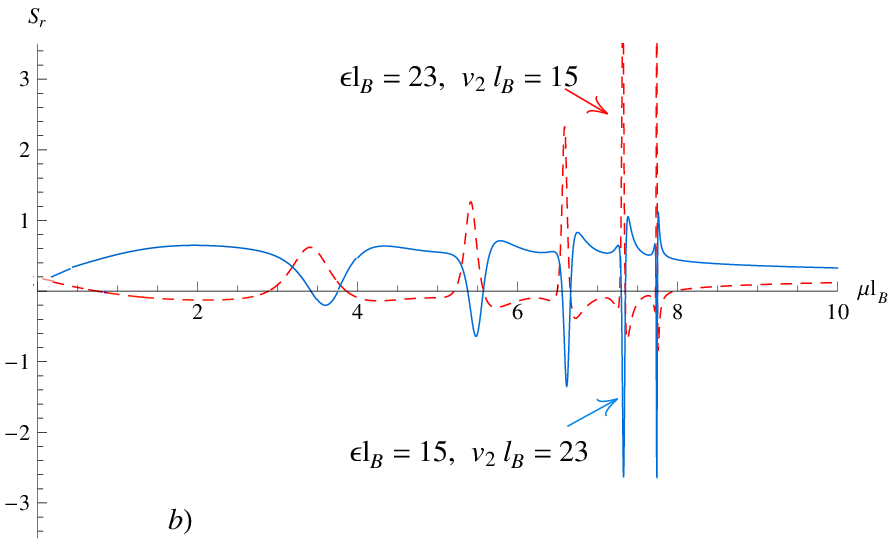}\\
\includegraphics[width=8cm, height=5cm]{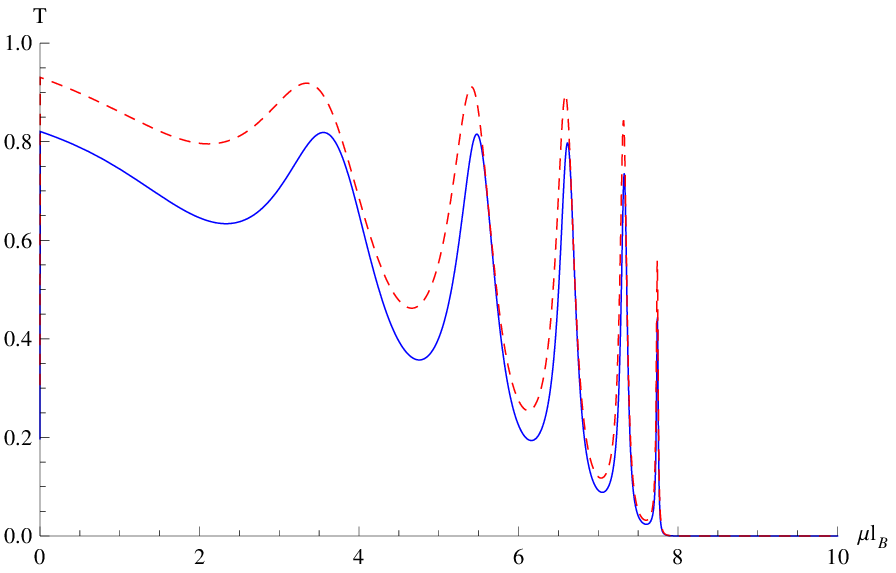}\ \ \ \
\includegraphics[width=8cm, height=5cm]{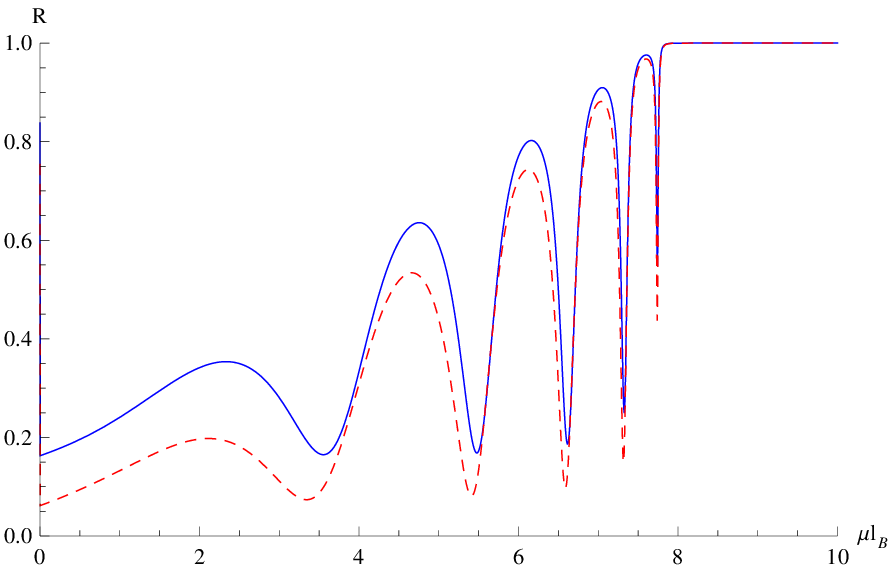}\\
%\end{figure}
%\begin{center}
 \caption{\sf{(Color online) a)/b): The GHL shifts $S_t/S_r$ and the probabilities $T/R$  versus the energy gap
$\mu l_{B}$ with   $\frac{d_{1}}{l_{B}}=1.1$,
 $\frac{d_{2}}{l_{B}}=1.5$, $k_{y} l_{B}=1$,
 $v_{1}l_{B}=25$, $(v_{2}l_{B}=23, \epsilon l_{B}=15)$,
 $(v_{2}l_{B}=15, \epsilon l_{B}=23)$.}}\lb{GHfig5}
\end{figure}

\begin{figure}[!ht]
\centering
\includegraphics[width=8cm, height=5cm]{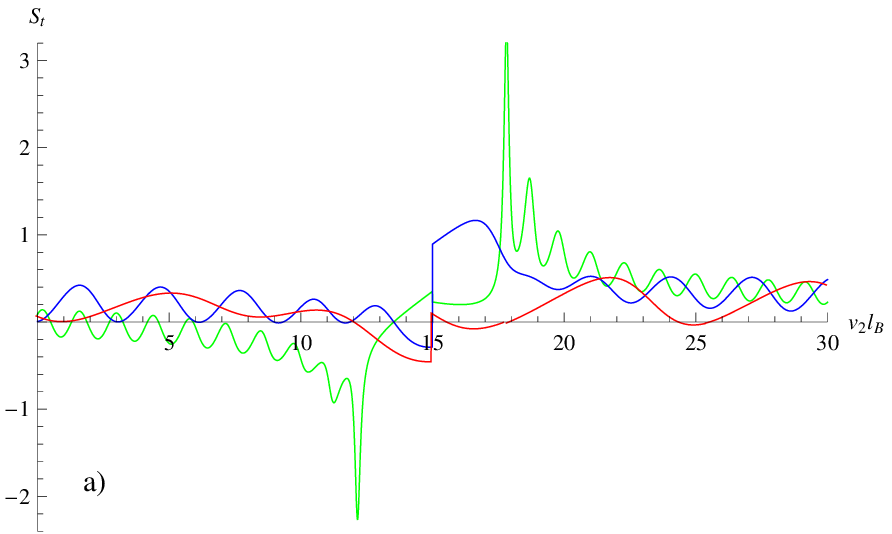}\ \ \ \
\includegraphics[width=8cm, height=5cm]{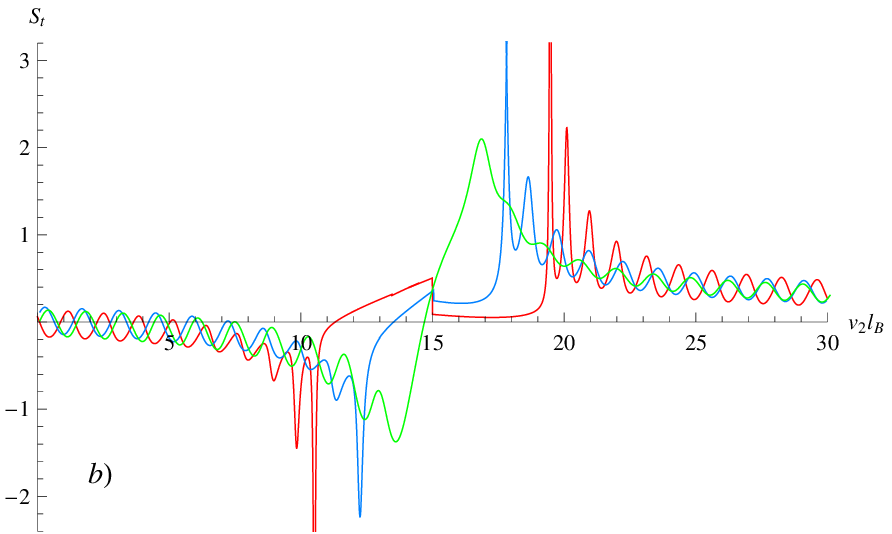}\\
%\end{figure}
%\begin{center}
 \caption{\sf{(Color online) The GH shifts $S_t$  versus the second potential height
$v_{2}l_{B}$ with $\frac{d_{2}}{l_{B}}=1.5$, $k_{y} l_{B}=1$,
 $v_{1}l_{B}=25$, $\epsilon l_{B}=15$. a): $\mu l_{B}=2$ and three different values
 of the barrier width  $\frac{d_{1}}{l_{B}}=1.1$ green color,
$\frac{d_{1}}{l_{B}}=0.5$ blue color, $\frac{d_{1}}{l_{B}}=0.2$
red color. b):
 $\frac{d_{1}}{l_{B}}=1.1$ and three different values of the gap
 $\mu l_{B}=0$ green color, $\mu l_{B}=2$ blue color,  $\mu l_{B}=4$ red color.}}\lb{GHfig6}
\end{figure}

Now let us investigate how the GH
shifts behave as function of the barrier potential  height $v_2 l_B$
which shown numerically in Figure \ref{GHfig6}a) for
different choices of the barriers width
$\frac{d_{1}}{l_{B}}=\{1.1,0.5, 0.2\}$ and in Figure \ref{GHfig6}b)
for different values of the energy gap $\mu l_{B}=\{0, 2,
4\}$.
It is clearly seen that the GHL shifts change their
sign near the point $v_2 l_B =\epsilon l_B$. We notice that when
the condition $v_2l_B >\epsilon l_B$ is fulfilled, the GH shifts
are positive, while it becomes negative when the height of the barrier
satisfies the condition $v_2l_B<\epsilon l_B$. We observe
from Figures \ref{GHfig6}a) and \ref{GHfig6}b) that the GH
shifts are strongly dependent on the barrier height $v_2l_B$,
which can be experimentally implemented by applying a local top
gate voltage $v_2l_B$ to graphene \cite{MMekkaoui}. This tells us that the GH
shifts can be controlled by changing the potential height $v_2l_B$.

\begin{figure}[!ht]
\centering
\includegraphics[width=8cm, height=5cm]{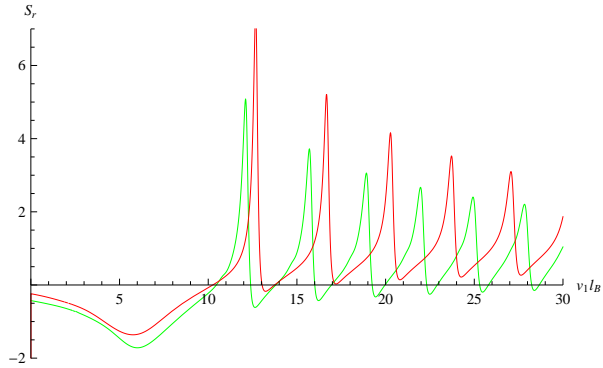}\\
%\end{figure}
%\begin{center}
 \caption{\sf{(Color online) The GH shifts $S_r$  versus the first potential height
$v_{1}l_{B}$ with   $\frac{d_{1}}{l_{B}}=0.2$ green color,
$\frac{d_{1}}{l_{B}}=0.5$ red color,
 $\frac{d_{2}}{l_{B}}=2.5$, $k_{y} l_{B}=1$,
 $v_{2}l_{B}=10$, $\epsilon l_{B}=7$,
 $\mu l_{B}=4$.}}\lb{GHfig7}
\end{figure}

   Figures \ref{GHfig7} shows the GH shifts in refection $S_r$ through a graphene
   triangular double barriers with
$\frac{d_{2}}{l_{B}}=2.5$, $k_{y} l_{B}=1$,
 $v_{2}l_{B}=10$, $\epsilon l_{B}=7$,  $\mu l_{B}=4$  and different values of the barrier
 width $\frac{d_{1}}{l_{B}}=\{0.2, 0.5\}$.  The
GHL shifts $S_r$  can be either negative or positive  and can
also be modulated for potential barrier $v_{1}l_{B}>v_{2}l_{B}$.

{Figures \ref{GHfig9}  illustrates the  dependence of the GH
shift and transmission probability  on the magnetic field $B$. Fo this, we 
choose three values 
$B=0.2 T$ red color, $B=1.5 T$ green color, $B=3 T$ blue color and
the other physical parameters are, respectively, $\mu=2$,
$d_{1}=1.1$, $d_{2}=1.5$, $k_{y}=1$, $v_{1}=25$. As shown in
Figure \ref{GHfig9}a), we plot the GH shifts in transmission
$S_{t}$ and the transmission probability $T$ as a function of
energy $\epsilon$ with $v_{2}=15$ and  found that the GH shifts
display sharp peaks inside the transmission gap. The GH shifts can
be changed from positive to negative by controlling the strength
of the magnetic field. However, the GH shifts finally become
negative with increasing the strength of the magnetic fields. We
can see that the transmission decreased with the increased
magnetic fields. In Figures \ref{GHfig9}b) we plot the GH shifts in
transmission $S_{t}$ and the transmission probability $T$ as a
function of $v_{2}$ with $\epsilon=15$. The GH shifts change sign
at the Dirac point $v_{2}=\epsilon$. In particular the GH shifts
in transmission can be negative and positive also the transmission
probability decreased with the increased magnetic fields.}

\begin{figure}[!ht]
\centering
\includegraphics[width=8cm, height=5cm]{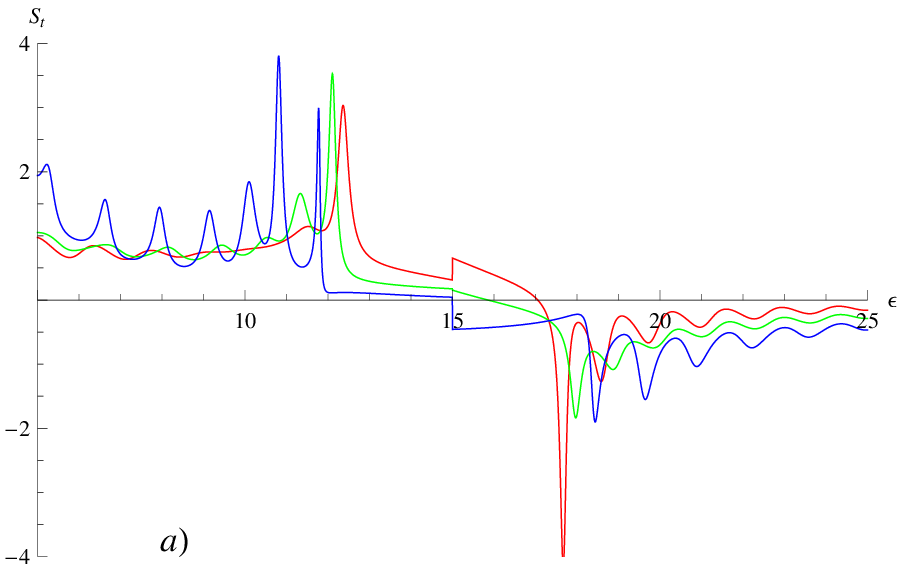}\ \ \ \
\includegraphics[width=8cm, height=5cm]{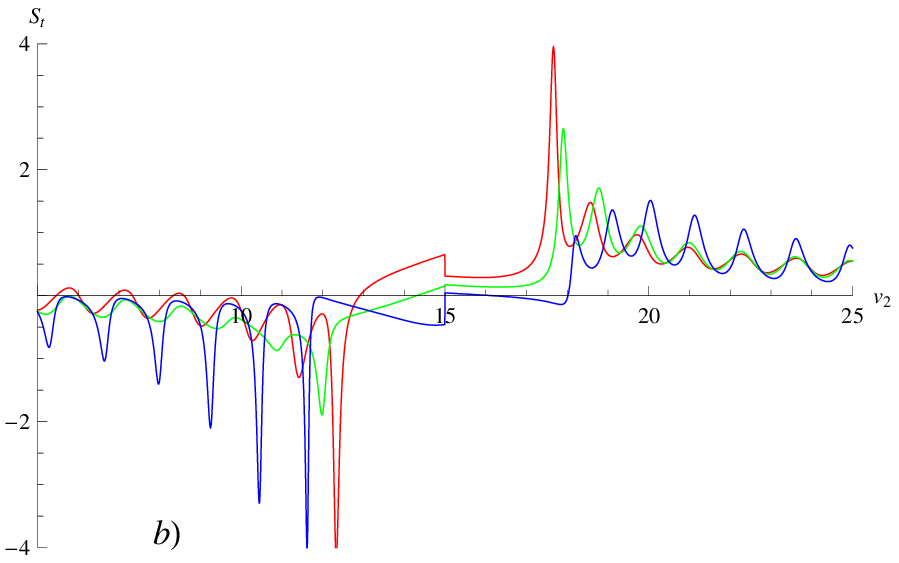}\\
\includegraphics[width=8cm, height=5cm]{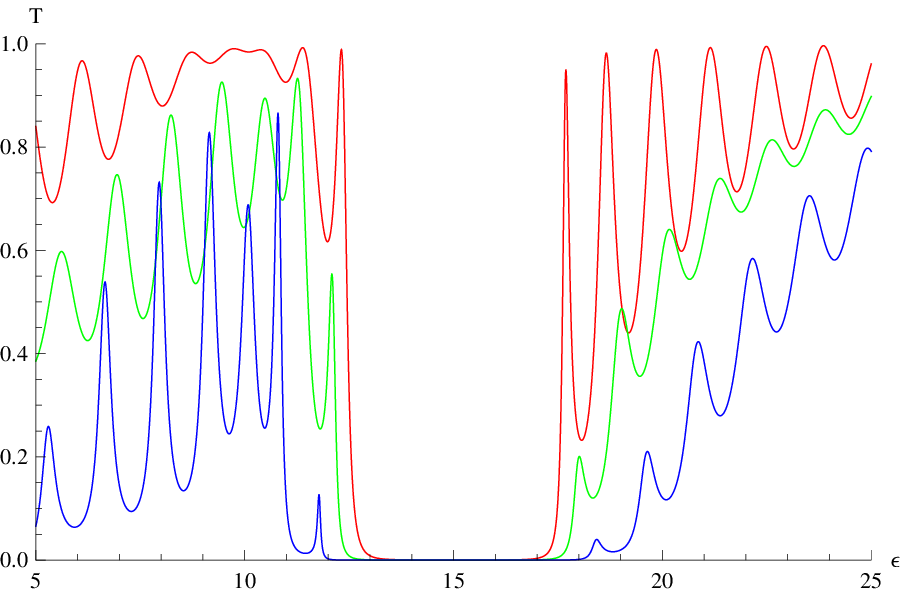}\ \ \ \
\includegraphics[width=8cm, height=5cm]{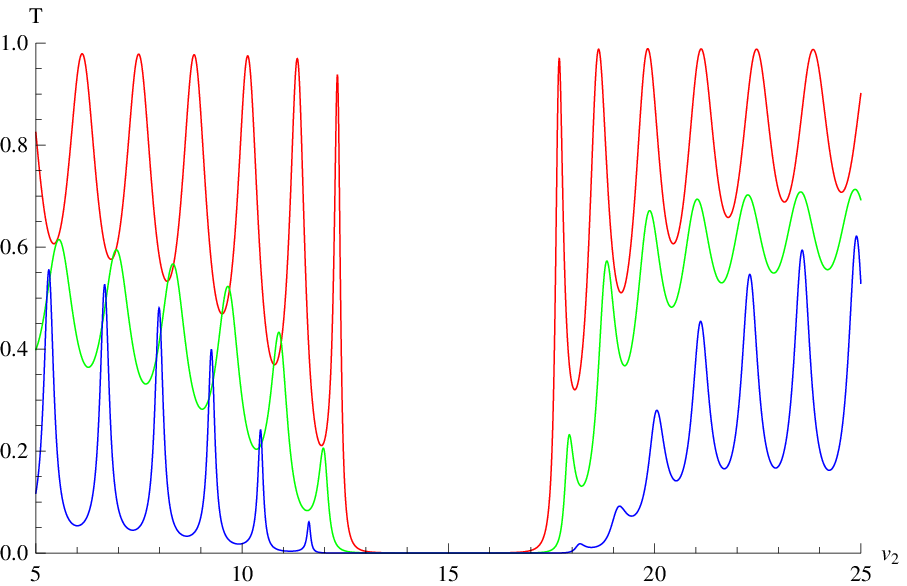}\\
%\end{figure}
%\begin{center}
 \caption{\sf{(Color online) a)/b) The GHL shifts $S_t$ and the probability transmission $T$ b) versus the energy
 $\epsilon/v_{2}$ for three different values
 of the magnetic field $B=0.2 T$ red color, $B=1.5 T$ green color, $B=3 T$ blue color,
 with $\mu=2$, $d_{1}=1.1$,
 $d_{2}=1.5$, $k_{y}=1$,
 $v_{1}=25$,  $v_{2}=15/\epsilon=15$.}}\lb{GHfig9}
\end{figure}

%%%%%%%%%%%%%%%%%%%%%%%%%%%%%%%%%%%%%%%%%%%%%%%%%%%%%%%%%%%%%
\section{Conclusion}
%%%%%%%%%%%%%%%%%%%%%%%%%%%%%%%%%%%%%%%%%%%%%%%%%%%%%%%%%%%%%%%

We have investigated the Goos-H\"anchen (GH) shifts for Dirac fermions
in graphene
scattered by
triangular double barriers and in the presence of a uniform magnetic
field.
%In doing so, we have set the materials needed to
%analytically determine and numerically analyze the GH shifts
%in transmission and in reflection.
In the first stage, we have solved
%These have been done by solving
the eigenvalue equation to end up
with the solutions of the energy spectrum in terms of different
physical parameters characterizing the five regions composing
the present system.

%in the Hamiltonian system.
In the second stage, we have used
%Using
the continuity conditions at each interface of the five regions
and used the transfer matrix  method to explicitly determine
the transmission and reflection coefficients. To make a link with
optics system, we have mapped our eigenspinor solutions
as the incident, reflected and transmitted beams together with a Gaussian function.
Subsequently, we have written
the obtained coefficients  in complex notation
to get  the transmission and reflection angles. These
were used together with the transverse wave vector $k_{y}$
around a point $k_{y0}$
%under some conditions
to derive the corresponding
GH shifts in transmission and in reflection amplitudes.

Different numerical results were presented in terms of
the physical parameters characterizing the present system. In fact,
Our
results show that the GH shifts are affected by the internal
structure of the triangular double barriers. In particular, the GH
shifts change sign at the transmission zero energies and peaks at
each bound state associated with the triangular double barriers.
It is observed that
%numerical results show that
the GH shifts can be enhanced by the
presence of resonant energies in the system when the incident
angle is less than the critical angle associated with total
reflection.

%%%%%%%%%%%%%%%%%%%%%%%%%%%%%%%
\section*{Acknowledgments}
%%%%%%%%%%%%%%%%%%%%%%%%%%%%%%

The generous support provided by the Saudi Center for Theoretical
Physics (SCTP) is highly appreciated by all authors. AJ
and HB
acknowledge the support of King Fahd University of Petroleum and
minerals under research group project RG171007.

\end{document}